# Essential nonlinearities in hearing

V. M. Eguíluz[a,b], M. Ospeck[c,d], Y. Choe[c], A. J. Hudspeth[c,d], and M. O. Magnasco[b],
[a]Instituto Mediterráneo de Estudios Avanzados IMEDEA (CSIC-UIB), E-07071 Palma de Mallorca (Spain)
Laboratories of [b]Mathematical Physics and [c]Sensory Neuroscience and [d]Howard Hughes Medical Institute
The Rockefeller University, 1230 York Avenue, New York, NY 10021
(March 16, 2000)

Our hearing organ, the cochlea, evidently poises itself at a Hopf bifurcation to maximize tuning and amplification. We show that in this condition several effects are expected to be generic: compression of the dynamic range, infinitely sharp tuning at zero input, and generation of combination tones. These effects are "essentially" nonlinear in that they become *more* marked the smaller the forcing: there is no audible sound soft enough not to evoke them. All the well-documented nonlinear aspects of hearing therefore appear to be consequences of the same underlying mechanism.

The classic Helmholtz theory [1] posits that our hearing organ, the cochlea, is arranged like a harp or the backplane of a piano, with a number of highly tuned elements arrayed along a frequency scale, performing Fourier analysis of the incoming sound. Although the notion that the inner ear works like a musical instrument offers a beautiful esthetic symmetry, it has serious flaws. In the 1940s, Gold [2] pointed out that the cochlea's narrow passageways are filled with fluid, which dampens any hope of simple mechanical tuning. He argued that the ear cannot operate as a passive sensor, but that additional energy must be put into the system. As in the operation of a regenerative receiver [3], active amplification of the signal can compensate for damping in order to provide highly tuned responses.

Von Békésy's classic measurements in the cochlea [4] demonstrated the mapping of sound frequencies to positions along the cochlea. He observed the tuning to be quite shallow and found cochlear responses to behave linearly over the range of physiologically relevant sound intensities. Gold's notions were largely set aside in favor of the hypothesis of coarse mechanical tuning followed by a "second filter," whose nature was surmised to be electrical.

Von Békésy conducted his measurements on cadavers, whose dead cochleas lacked power sources or amplifiers that might have provided positive feedback. Only fairly recently, laser-interferometric velocimetry performed on live and reasonably intact cochleas has led to a very different picture [5,6]. There is, in fact, sharp mechanical tuning, but it is *essentially nonlinear*: there is no audible sound soft enough that the cochlear response is linear. Although the response far from the resonance's center is linear, at the resonance's peak the response rises sublinearly, compressing almost 80 dB into about 20 dB (Fig. 1). The width of the resonance increases with increasing amplitude, being least for sounds near the threshold of hearing. Observation of the response's essential nonlinearity at the level of cochlear mechanics contradicts von Békésy's finding. Furthermore, this nonlinearity does not originate in the rigidity of membranes or in fluid-mechanical effects. Because it reversibly disappears if the cochlea's ionic gradient is temporarily disturbed, the nonlinearity depends on a biological power supply [7].

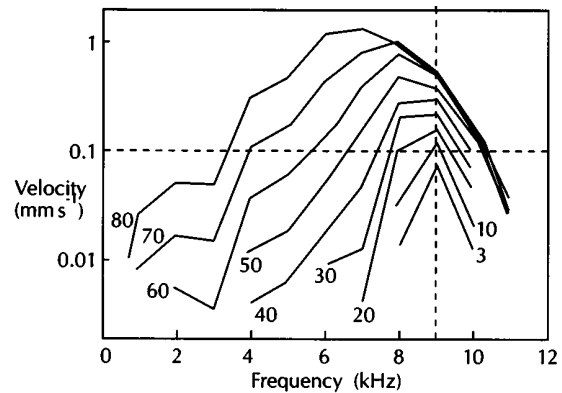

FIG. 1. Laser velocimetric data from a living chinchilla's cochlea displaying the root-mean-square velocity of one point on the basilar membrane as a function of driving frequency. Each curve represents a different level of stimulation, labelled in decibels sound-pressure level. The characteristic frequency at the position of measurement is 9 kHz. Notice that at 4 kHz, the curves from 40 dB to 80 dB span two decades (40 dB), whereas at 9 kHz the curves from 3 dB to 80 dB span just under one decade (20 dB). Note that the response at 9 kHz saturates beyond 60 dB. At 4 kHz, the response rises an average of 1 dB per decibel, whereas at 9 kHz the response rises only 0.3 dB per decibel. Note furthermore the dramatic increase in bandwidth as the intensity increases. Courtesy of M. A. Ruggero [5].

Gold conjectured that a regenerative mechanism for hearing could lead to feedback oscillations so that the ear would actually emit sound. The discovery that the ear indeed produces spontaneous otoacoustic emissions [8] rekindled interest in Gold's theory. Recently these emissions have been found to be limit cycle oscillations [9]. The perceived pitch of missing fundamental tones has additionally been shown to be compatible with a relaxation oscillation locking mechanism [10].

Psychoacoustical experiments have provided another means of probing the nonlinearities of hearing. When two sine waves traverse a system with a nonlinear transfer function, the response includes combination tones, integer linear combinations of the input frequencies, whose amplitudes scale as products of the input amplitudes raised to the appropriate positive integer powers. If the input is weak enough, a linear "small-amplitude" regime should be recovered in which combination tones are absent. Psychoacoustical experiments showed that the perceived intensity of combination tones is not suppressed in this fashion: although the $2f_1 - f_2$ combination tone should decline by 3 dB for each decibel of attenuation in the input sound, the actual attenuation is only 1 dB per decibel [11]. The intensity relative to the fundamental tones remains constant. These observations, too, imply that the system is essentially nonlinear: no audible sound is faint enough to elicit a small-amplitude, linear regime.

We shall show that all of these apparently disparate characteristics are related to one another, stemming from the same mechanism. In dynamical systems language, we would say that Gold's theory asserts that the elements of the hearing organ somehow *poise* themselves at a Hopf bifurcation, like a sound technician adjusting the volume at an amplifier to the loudest possible setting before feedback oscillation ensues. We shall show that at a Hopf bifurcation we generically expect essential nonlinearities, compression of dynamic range, sharp tuning for small input, and broad tuning for large input. In essence, several nonlinear aspects of hearing may stem from the Hopf bifurcation. We shall then argue that given our current understanding of hair-cell physiology it is plausible that this is occurring.

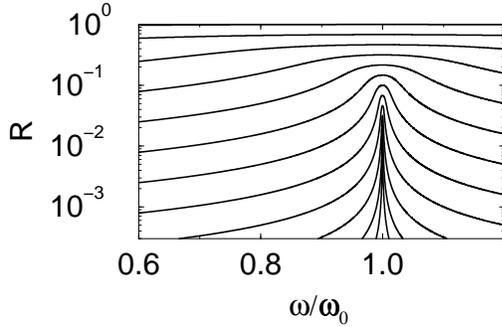

FIG. 2. Hopf resonance. The response $R$ to different levels of forcing $F$ is obtained from Eq. (2); the amplitude of forcing increases in increments of 10 dB for successive curves from bottom to top. At resonance the response increases as the one-third power of the forcing, whereas away from the resonance the response is linear in the forcing.

A generic equation describing a Hopf bifurcation can be written

$$\dot{z} = (\mu + i\omega_0)z - |z|^2 z$$

where $z(t)$ is a complex variable of time, $\omega_0$ is the natural frequency of oscillation, and $\mu$ is the control parameter [12]. When $\mu$ becomes positive, the solution $z \equiv 0$ becomes unstable, and a stable oscillatory solution appears, $z = \sqrt{\mu}\exp(i\omega_0 t)$. If the system is subjected to periodic forcing as $\dot{z} = (\mu + i\omega_0)z - |z|^2 z + Fe^{i\omega t}$, then for the spontaneously oscillating system a variety of well-studied entrainment behaviors occur. Assuming a 1:1 locked solution of the form $z = Re^{i\omega t + i\phi}$, we obtain

$$F^2 = R^6 - 2\mu R^4 + (\mu^2 + (\omega - \omega_0)^2)R^2 \qquad (1)$$

This equation is a cubic in $R^2$ and hence solvable:

$$3R^2 = S^{1/3} + 2\mu - U_2 S^{-1/3}$$

where

$$2S = D + \sqrt{D^2 + 4U_2{}^3} \qquad D = 27F^2 + 16\mu^3 - 18\mu U_1$$

$$U_1 = \mu^2 + (\omega_0 - \omega)^2 \qquad U_2 = -\mu^2 + 3(\omega_0 - \omega)^2$$

If we specialize Eq. (1) exactly at the bifurcation we obtain

$$F^2 = R^6 + (\omega - \omega_0)^2 R^2 \qquad (2)$$

from which we can demonstrate directly one of our main contentions. At the center of the resonance, where $\omega = \omega_0$, $R \approx F^{1/3}$, so no matter how small $F$ might be, the response is nonlinear (Fig. 2). Notice that because a cubic root of a small number is much larger than the number, the amplification $R/F$ or the differential amplification $dR/dF$ blows up as $F^{-2/3}$ for infinitesimal forcings. Away from the resonance's center, for sufficiently small $F$ we obtain $R \approx F/|\omega - \omega_0|$, the standard form for a single pole seen from a distance; for any $\omega$, the amplification is constant and independent of $F$.

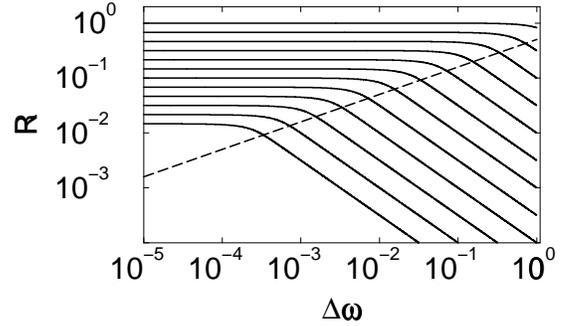

FIG. 3. The resonance of Fig. 2 in log-log form shows the compressive regime (to the upper left) and the linear regime (to the lower right) and the boundary between them. The dashed line given by Eq. (3) indicates the half-width $\Gamma$.

The definitions of "near the resonance" and "far from the resonance" depend on the amplitude of the forcing; therefore the interface between the two regimes depends on $F$. If we define the half-width $\Gamma$ of the resonance as the range in $\omega$ for which $R$ falls by one-half (Fig. 3), $(R/2)^6 + \Gamma^2(R/2)^2 = R^6$, from which

$$\Gamma = \frac{3\sqrt{7}}{4} F^{2/3} \qquad (3)$$

For this system the gain-bandwidth product is constant and independent of the forcing. The gain-bandwidth balance depends strongly on the forcing amplitude, however, asymptoting to infinite gain and zero bandwidth for zero forcing amplitude. This behavior strikingly resembles that of the velocimetric data for the basilar-membrane response [5].

As noted, if the control parameter lies exactly at the Hopf bifurcation, there is no forcing soft enough not to elicit a nonlinear response. This is no longer true if the parameter is not poised *exactly* at the bifurcation. *Near* the bifurcation, there is a linear regime for soft enough sounds; how soft they need to be depends upon closeness to the bifurcation. The precision with which the system can be so poised determines the maximal amplification and frequency selectivity. We again specialize Eq. (1), this time to the case $\omega = \omega_0$, exactly at resonance, to get

$$F^2 = R^2(R^2 - \mu)^2 \qquad (4)$$

Consider first $\mu < 0$, the sub-bifurcation regime. As $F \to 0$ then $R \to -F/\mu$: the amplification for infinitesimally soft sounds is $-1/\mu$, which becomes infinite only exactly at the transition. For $\mu$ sufficiently small and negative we observe compressive nonlinearity for $F > (-\mu)^{2/3}$ and a linear regime for softer sounds. We should furthermore note that for Eq. (1), $\mu$ is also the parameter for exponential relaxation in the absence of forcing: the system relaxes to the quiescent state as $\exp(\mu t)$. Thus the linear-regime amplification is exactly proportional to the integration time given by this relaxation; this integration time becomes infinite exactly at the bifurcation. Cochlear velocimetry data show the response becoming linear again below the hearing threshold [6]; this observation raises the possibility that the feedback loop controlling the poising operates through the very same signal used for detection.

Once past the Hopf bifurcation ($\mu > 0$), an oscillation occurs, for which the response above is by definition phase-locked 1:1, so its stability is constrained to the 1:1 Arnold tongue. In order to fully explore the behavior of the system around the Hopf bifurcation, it is better to consider the simplest forced model able to suffer quasiperiodic transitions. The best numerical scheme is to define a system whose solution we can compute analytically, then to force it impulsively so that we obtain a closed-form iterated map [13,14]. The simplest such homogeneous oscillator is

$$\dot{r} = r(\mu - r^2) \qquad (5)$$
$$\dot{\theta} = \omega \qquad (6)$$

Numerical exploration of this model and of the model described in [12] shows that the features described above are independent of model details (V. M. Eguíluz, in preparation).

We have thus established that several nonlinear aspects of hearing are compatible with the idea that the cochlea poises itself at a Hopf transition. How might this behavior originate in the hearing organ? One possibility is that the response dynamics of individual sound-sensing elements—the hair cells of the inner ear—itself displays a Hopf bifurcation. We shall next examine physiological evidence in support of this proposition.

Individual hair cells show electrical frequency selectivity, being tuned to specific frequencies by resonance of the membrane potential [15]. A seven-dimensional conductance-based model describes the hair cell's electrical amplifier, called the membrane oscillator [16]. In this model, the hair cell's capacitance is charged by current through the transduction channels, then discharged by $Ca^{2+}$-activated $K^+$ current. The model's control parameter $\mu$ is a strong function of both the transduction and the $Ca^{2+}$ conductances. As described by the membrane-oscillator model with increased $\mu$, electrically resonant hair cells in the hearing organs of amphibians, reptiles, and birds operate near a supercritical Hopf bifurcation. A small conductance oscillation in the transduction channels engenders a large current-to-voltage gain, the benefit of operating near a Hopf bifurcation (M. Ospeck *et al.*, in preparation).

In lower vertebrates, frequency-specific amplification in the auditory system derives in part from mechanical properties of the hair bundle, the mechanoreceptive organelle of the inner ear [17,18]. This bundle does not behave as a merely passive transducer. Evoked hair-bundle oscillations instead demonstrate that the hair bundle is capable of producing active transient motions and of amplifying mechanical inputs [21–24]. Moreover, hair bundles can produce limit-cycle oscillations, a phenomenon that may underly otoacoustic emissions [25]. Finally, a hair bundle can generate combination tones similar to those found in psychoacoustical experiments [26].

Two suggestions have been made about the mechanism of hair-bundle oscillations [17]. Both posit that the force-generating elements regulate the elastic properties of the mechanoelectrical transduction channel in a $Ca^{2+}$-dependent manner, thus modulating tension in the associated gating spring and altering the mechanics of the hair bundle. One possibility is that myosin molecules anchoring the channel complex to the actin core of the stereocilia power the oscillations. The alternative proposal is that the channel complex itself is intrinsically active and generates force. The primary supposition of this model [27] is that the closed state of the channel is stabilized by $Ca^{2+}$ binding. Because there is a $Ca^{2+}$ concentration gradient across the cell membrane and the channel is permeable to $Ca^{2+}$, channel opening regulates the local intracellular $Ca^{2+}$ concentration and thus the force generated through channel reclosure.

Variation of the model's parameter values through a physiologically plausible range reveals a locus of Hopf bifurcations whose frequencies span the range of human hearing. Near the bifurcation, one observes compressive frequency selectivity; the system is essentially nonlinear. One particularly relevant control parameter is the number of stereocilia in the hair bundle: many of the mechanical properties may be de-fined as functions of this value, which is clearly regulated along the cochlea. In agreement with experiment, near the bifurcation locus the model maps tall, thin hair bundles to the low-frequency range and short, broad bundles to higher frequencies. A second control parameter governs $Ca^{2+}$-binding kinetics; faster transitions correspond to higher oscillation frequencies. Tuning of this parameter may be achieved through modulation of the intracellular $Ca^{2+}$ concentration, which is also subject to tight regulation [28]. These models demonstrate that hair cells can operate near a Hopf bifurcation for realistic parameter values.

We have shown that tuning to a Hopf bifurcation can account for three well-documented essential nonlinearies of the ear: compression of dynamic range, sharper cochlear tuning for softer sounds, and generation of combination tones. The great advantage of the regenerative tuning strategy is that it requires a minimal number of active elements; because the tuner and the amplifier are one and the same, this mechanism is evolutionarily accessible. We have also reviewed evidence that the sensory receptors of the cochlea, the hair cells, operate near a Hopf bifurcation. Because the cochlea is a complex geometrical structure traversed by nonlinear waves, relating the contribution of individual hair cells to the behavior of the entire organ remains a both a theoretical and an experimental challenge. It is important to determine, for instance, whether hair bundles are stiff enough to affect the propagation of the cochlear traveling wave and whether the hair cells' electrical responses affect hair-bundle movement. Despite the difficulty in linking the ear's microscopic to its macroscopic behavior, though, it seems likely that the process that poises the ear as a whole near a Hopf bifurcation is identical to that involved in bringing each hair cell to a bifurcation. A hair cell in the cochlea can measure its input only in the context of the organ as a whole, and the only active element positioned to adjust the cochlea's behavior is the hair cell.

We thank Albert Libchaber, Oreste Piro, Mario Ruggero and Bruce Knight for enlightening discussions. After making our initial submission, we received a manuscript by S. Camalet, T. Duke, F. Jülicher, and J. Prost describing a model for self-tuned critical oscillations of hair cells; we thank those authors for communicating their unpublished observations. We acknowledge support of the Mathers Foundation and an EEC Visiting Fellowship for VME. Part of this work was supported by National Institutes of Health grant DC00241. YC is supported in part by an NSF Graduate Fellowship. MO is an Associate and AJH an Investigator of Howard Hughes Medical Institute.